# Complex physical properties of an adaptive, self-organizing biological system

József Prechl

R&D Laboratory, Diagnosticum Zrt., Budapest H-1047, Hungary

e-mail: jprechl@diagnosticum.hu

**Abstract**

The physical interpretation of the functioning of the adaptive immune system, which has been thoroughly characterized on genetic and molecular levels, provides a unique opportunity to define an adaptive self-organizing biological system in its entirety.

Here I describe a configuration space model of immune function, where directed chemical potentials of the system constitute a space of interactions. In the physical sense, the humoral adaptive immune system adjusts the chemical potential of all available antigenic molecules by tuning the chemical potential and organizing the network hierarchy of its sensor-effector molecules, antibodies. The coupling of sensors and effectors allows the system to adjust the thermodynamic activity of antigens and antibodies, while network organization helps minimize chemical potentials and maximize diversity. Mathematically the system couples the variance of Gaussian distributed interaction energies in its interaction space to the exponentially distributed chemical potentials of its effector molecules to maintain its stationary state. This process creates a scale-free network in interaction space, where absolute thermodynamic activity corresponds to node degree. In the thermodynamic interpretation, the system is an ensemble carrying out $\mu N$ work, adjusting chemical potentials according to the changes in the chemical potentials of the surroundings. The validity of the model is supported by identifying an interaction flexibility index, the corresponding variables in thermodynamics and network science, and by confirming its applicability to the humoral immune system.

Overall, this statistical thermodynamics model of adaptive immunity describes how adaptive biological self-organization arises from the maintenance of a scale-free, directed interaction network with fractal topology.

# 1. Introduction

Macromolecules, and in particular proteins, have evolved as part and basis of the evolution of life. The diversity of proteins reflects the diversity of life itself, while the tree of evolution of proteins reflects evolution of life (1). Proteins serve to build (structural proteins) and operate (enzymes, transporters, regulators, secretions) an organism. For all these functions it is necessary for proteins to interact with other molecules, from metal ions through small molecules to other macromolecules. Beyond spatial arrangement, that is, the necessity of being at the same place at the same time, interactions require thermodynamic probability for the binding. In other words, the system searches for a free energy minimum state while proteins sample interacting partners. This in turn means that most proteins have energy minimums in a bound state. The evolution of proteins from this aspect is the search for binding partners: genetic changes resulting in new, modified proteins will be sustained if the binding is advantageous for the survival of the organism. Since paired interactions create networks of interactions, evolutionary processes are constrained by energy transduction networks (2). These biological networks may mediate metabolic fluxes (3), protein interactions (4) or information in neural networks (5).

The functioning of the adaptive immune system is based on directed evolution of proteins, called antigen receptors, of lymphocytes (6,7). In its essence, the adaptive immune system is a catabolic system: it removes molecules and cells from the body, based on molecular immunological instructions. Humoral adaptive immunity utilizes glycoproteins called antibodies for tagging targets for removal (8). Antibodies, unlike all other somatic proteins, are not encoded in the genome but are produced as a result of genetic recombination and mutation events during our lifetime (sometimes referred to as accelerated evolution). The role of the immune system is to drive and direct the evolution of these molecules so as to maintain molecular and cellular integrity of the host: this is what we call immunity. As a result, millions of distinct antibodies are produced constantly in our bodies, removing cellular debris, maintaining balance with the microbiota and fending off invading agents (9).

On the molecular level these immunological events have been characterized in detail, however the systems level understanding and physical description of self-organization of the adaptive immune system is still incomplete. Since the adaptive immune system is a *bona fide* complex adaptive system considering the number of its elements, its diversity and self-organizing capability, this bigger picture can be approached by characterizing immunological phenomena using the terminology of physics of complex systems and mathematical models. Following upon the pioneering work of Perelson (10–12), the application of physical and mathematical models for the explanation of questions about immune repertoire size and diversity, lymphocyte population dynamics, immunological memory has regained interest (13–17). Here, I attempt to combine chemical thermodynamics with network theory, building up a model that is consistent with a novel technological approach of serum antibody reactivity measurement (18). The basic concept of this model was introduced in previous papers that showed how stages of B-cell differentiation correspond to the generation of a thermodynamic system and a network of interactions (19,20). In this paper we shall first define the biological system as a configuration space, using physical chemistry, in order to identify analogues of cells and molecules in a physical system. Then we shall use the mathematical description of the system to derive the architecture of the hierarchical network that governs antigen transport in a stationary state of the system. Finally, we discuss the model from the perspective of immunology, physical chemistry, network science and physics of complex systems.

## 2. Results and discussion

### 2.1. Compartments of configuration space

Throughout the article we shall examine the adaptive immune system in an abstract space called configuration space. Configuration here refers to the arrangement of atoms that contribute to the binding of the antibody, corresponding to a particular conformation of the binding surface, and the arrangement of all the constituent antibodies in the system. Since binding, the formation of non-covalent bonds between interaction partners, requires shape congruency, arrangement in this space determines both antibody and antigen structure, and binding specificity and strength. The manifold enclosing components of the immune system in configuration space thus defines both antibodies and antigens. In this space cells and molecules of the immune system, corresponding to clones bearing or being a particular antibody, respectively, are positioned according to their potential to interact with a target and in the direction of the target. The interaction potential is the chemical potential of the antibody in body fluids. Targets are antigens, molecules that drive the evolution of the system. Considering a three-dimensional Euclidean space as the configuration space of the system, target antigen shapes – epitope surfaces - form a continuity on an imaginary spherical canvas enclosing the system (Fig.1).

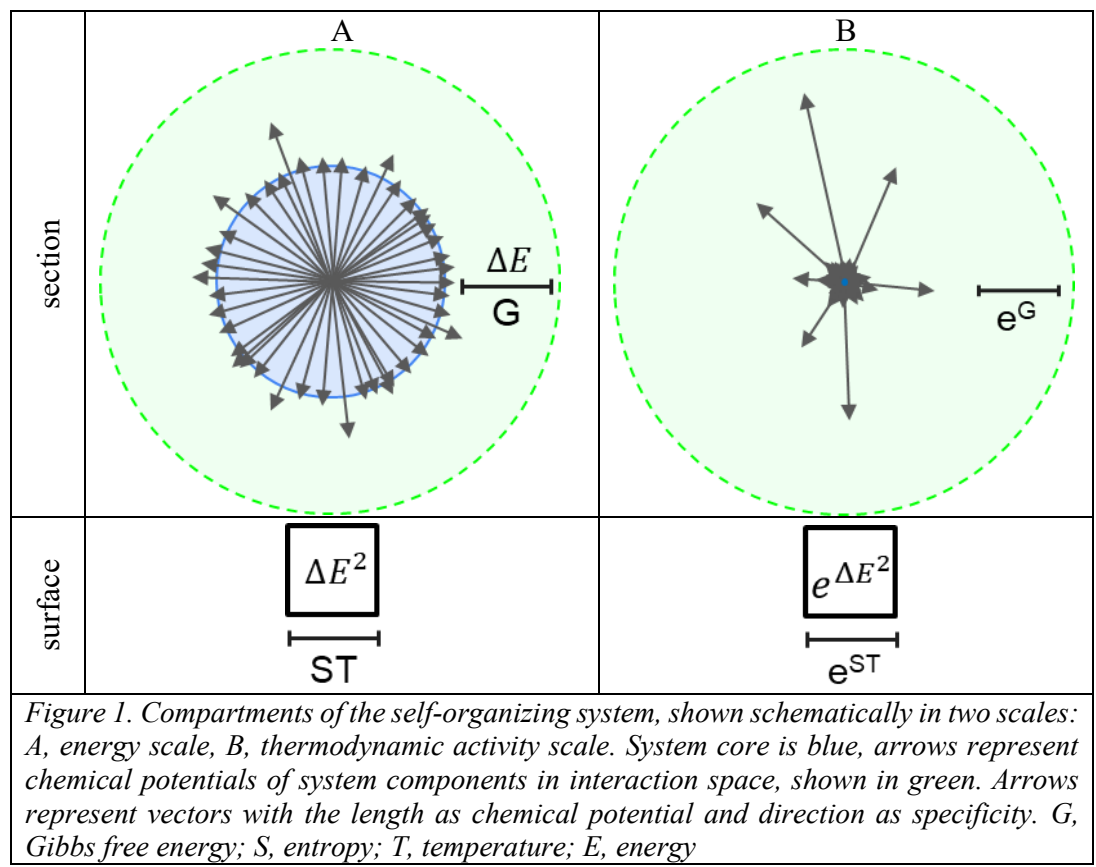

*Figure 1. Compartments of the self-organizing system, shown schematically in two scales: A, energy scale, B, thermodynamic activity scale. System core is blue, arrows represent chemical potentials of system components in interaction space, shown in green. Arrows represent vectors with the length as chemical potential and direction as specificity. G, Gibbs free energy; S, entropy; T, temperature; E, energy*

From the physical point of view the system is embedded in a reservoir, which serves as a heat and particle bath; this is the host organism that maintains constant temperature and pressure of the system. The system is in stationary state in the sense that it is constantly dispersing energy while maintaining its state functions. Overall, we can regard the system as an ensemble of overlapping and hierarchically arranged thermodynamic ensembles.

We consider three compartments in our configuration space model: 1, the core of the system, surrounded by the core surface; 2, interaction space around the core, extending from the core surface to the system surface, and 3, the system environment (surroundings) beyond system surface. We shall use two scales for examining and visualizing the system: energy scale (Fig.1A), which is the logarithmic form of the interaction probability or thermodynamic activity scale (Fig.1B). The system core contains the conceptual energy source of the system that is located in the origin of configuration space. Events within the core therefore can be thermodynamically unfavorable, consuming energy to maintain a continuous supply of particles in interaction space. The core surface is spherical, and is characterized by system components with average molecular free energy $\mu_0$.

The system is composed of two kinds of constituents. Cells, which carry sequence information, are treated as particles or nodes in the system. Soluble antibodies, which are energy transfer units connecting cells, are treated as links. Energy corresponds to antigen molecules, which are transferred from lower to higher affinity antibodies, with corresponding decrease in the system free energy. These constituents are present both in the core and in the interaction space. Interaction space is defined and confined by the totality of chemical potential vectors, each of which represents the interaction energy of a particular antibody clonotype. Importantly, cells and soluble antibodies belong to these clonal groups, determined by sequence and identified by the structure of the antibodies. Radial geometrical distance in this configuration space is measured in units of chemical energy per molecule, while vector directions identify structures, molecular shapes in the surroundings.

From the network point of view, in the humoral adaptive immune system nodes are B cells (sequence information, clonotype) at various stages of differentiation, while links are structural similarities between B cell clonotypes and thereby represent pathways of antigen transport. Physically it is the soluble circulating antibodies that transport antigen, and thereby delineate pathways connecting clonotypes.

**2.2. Principle of self-organization**

Self-organization is the ability of a system to arrange its constituents so as to adjust to its environment and also maintain a responsive state. Self-organizing systems tend to reach an optimal state associated with minimal interaction and dissipation (21)(22)(23). The immune system adjusts the quality (intensive physical property) and number of molecules (extensive physical property) in the host by the coupling of a sensor and an effector mechanism (20). The sensor mechanism is a cell that survives only in the presence of signals triggered by the target molecule via an antibody displayed on the cell surface. The effector mechanism is the generation of molecules, soluble antibodies, that bind both to the target molecule and to cells that remove the resulting complexes. Coupling means the balanced adjustment of the chemical potential $\mu$ of the antibody on the sensor surface and of the effector molecules against that molecule species. Biologically it means that the sensor and effector B cells are genetically closely, clonally related, expressing identical or similar antibodies either on their surface or as secreted components. Relations in terms of network connectivity are determined not only by similarity, but also by the similarity of interaction partners (24). In the case of the humoral immune system this means, that antibodies that are clonally not or only distantly related may contribute to the binding and elimination of a given antigen molecule. In other words, in a subnetwork of the system dedicated for a given antigen, distinct clonotypes may be linked together in the network of antigen transfer pathways. The concept of clone sizes and corresponding Ab chemical potential and antigen thermodynamic activity, with corresponding chemical potentials, can be visualized by activity maps and energy diagrams (Fig. 2). The thermodynamic activity of a particular antibody clone is related to the frequency of the cells belonging to that clone (clone size). The energy of an antibody is the logarithm of its activity. Similarity is represented by the topology of the activity map, with closely related clones being neighbors. The thermodynamic activity of an antigen is related to its ability to stimulate the immune system and the corresponding energy is its logarithm. Again, structural similarity is represented by location in the activity map.

The effector cells (plasma cells) secrete antibody, which bind antigen with an efficiency determined by chemical potential and antibody concentration. Following an active phase of the immune response a resting phase is established, along with the generation and maintenance of memory cells: these are the long-lived sensors and effectors of the system. We assume that in the resting phase of the immune response a steady state is maintained regarding antibodies and antigens. During the active phase of the immune response the chemical potential of the effector molecule is raised, more bound antigen will be removed from the system. At the same time the coupled increase in the sensor sensitivity will allow the cell to survive in the presence of less antigen, and steady state with higher chemical potential is established. Several different antibodies with distinct chemical potentials may contribute to antigen free energy adjustment, the system selects clones with the chemical potential and concentration required. Our argument here is that this selection process is governed by thermodynamic rules.

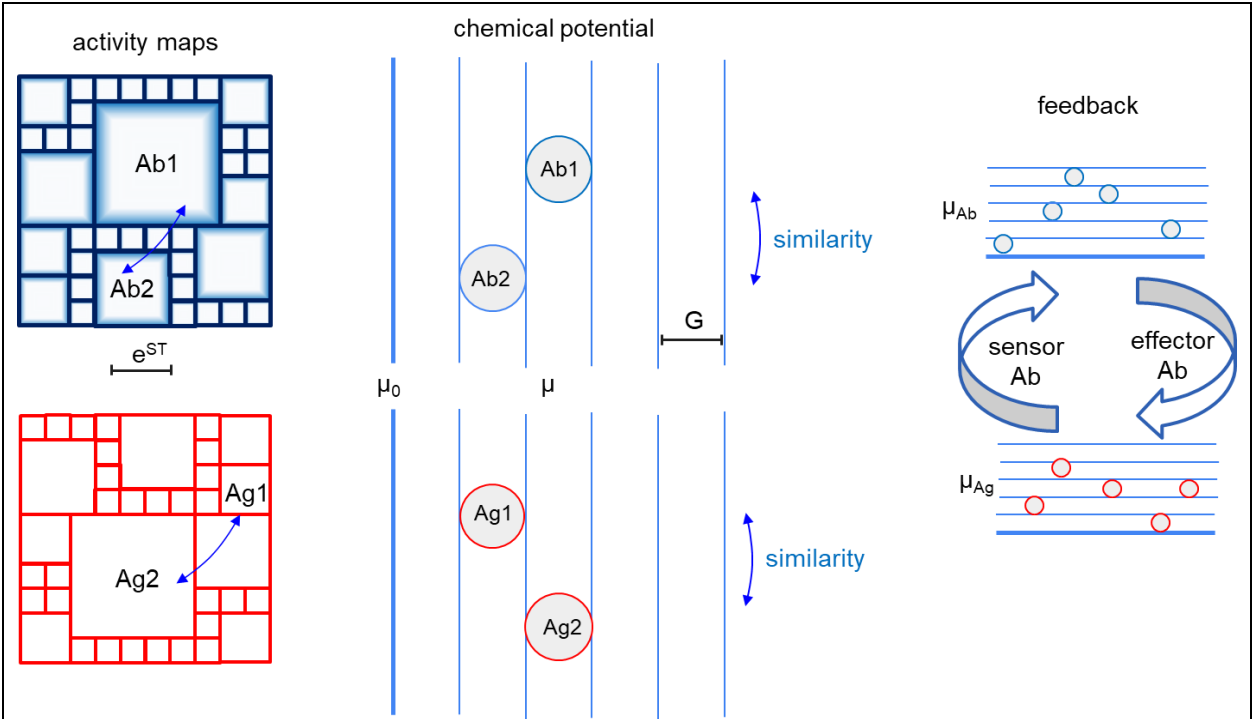

*Figure 2. Activity and energy diagrams of antibodies and antigens.*
*Thermodynamic activity and structural similarity appear in activity maps and energy diagrams for antibody (blue) and antigen (red). Square areas in the map represent activity, neighborhood distance (blue double arrows) means similarity. Continuous feedback mechanisms maintain stationary state.*

The cells of the adaptive immune system, a tissue that penetrates the whole organism, possesses a unique property no other cells have: lymphocytes are capable of directing the evolution of proteins within the organism (15). These proteins, called antigen receptors, are generated by genetic recombination and mutation and selection events, resulting in a diversity that well exceeds the diversity of all other proteins in the organism. Depending on the size of the host, lymphocyte numbers are in the range of $10^8$-$10^{12}$ cells (25,26). Generation of diversity is a random molecular process, while the function of the adaptive immune system is to adjust, by cycles of division and checkpoints of selection, lymphocyte specificity and affinity. These two features correspond to direction and length of vector in configuration space. A stationary state, which is immunological rest in our system, is reached by the adjustment of an intensive and an extensive property. The intensive property is the chemical potential of antibody, determined by cell's genes and antibody sequence and structure. The extensive property is diversity and numerosity of antibodies. Thus, in a broader sense, balancing is between molar free energy and system entropy.

Self-organization also maintains responsiveness in the system. In our case it is maintained by the sensors, B cells displaying membrane-bound Ab, that are able to initiate an immune response and thereby reset the chemical potential of both sensors and effectors and lead to the reorganization of local hierarchy in the system. Activity maps and energy diagrams in this paper therefore represent a particular point in the lifetime of a highly dynamic system.

## 2.3. Architecture of system

The immune system is theoretically capable of interacting with any molecular structure. That means it takes up volume in all directions of antigen space, creating a spherical core (Fig.3). It has to allocate its limited resources for the generation of particles in interaction space in the most energy efficient way. Therefore, it is split up into a number of thermodynamic ensembles, distributing its resources in-between these. A particular ensemble is directed against a particular target epitope, adjusting its properties to attain stationary state. Because resources of the system are finite, ensembles compete for them and may overlap, cluster and form hierarchies. At any one moment in time the distribution of the state of the ensembles is the distribution of chemical potentials and their way of sharing system energy. The configuration space of the self-organizing system is a systemic ensemble in the sense that it is a collection of coexisting grand canonical ensembles, each ensemble responding to a different antigenic component of the environment. In the vector space of chemical potentials of the system, the location of each ensemble identifies a particular direction, representing the potential energy of interaction with the particular part of the environment.

The probability density function of absolute thermodynamic activity $\lambda$ of entities of the system, as determined by chemical potentials $\mu_{Ab} \sim \ln(\lambda_{Ab})$, can be modeled by a mixture of exponential and lognormal distributions, as proposed in the RADARS model (19). The model is based on the assumptions that 1, chemical potentials of antigens in interaction space, corresponding to free energy of interactions of the system with the surroundings, are distributed exponentially, similar to a Boltzmann distribution, 2, equilibrium binding constants of antibodies are distributed lognormally (27), with a corresponding normal distribution of molar free energies of binding, and 3, the immune system arranges and adjusts antibody chemical potentials to maintain steady state.

The mixture of exponential and lognormal distributions was described by Reed (28) and Mitzenmacher (29), which we use here with the following parametrization:

$$p(\lambda) = \int_0^\infty \alpha e^{-\alpha \mu} \frac{1}{\sqrt{2\pi\mu}} \frac{1}{\lambda} e^{-\frac{1}{2}\frac{(\ln\lambda)^2}{\mu}} d\mu \qquad (1)$$

where
$\alpha$ is the rate of the exponential distribution of chemical potential of antigen,
$\lambda$ is the absolute thermodynamic activity of antibody,
$\mu$ is the chemical potential of antigen, the distance from system average chemical potential $\mu_0$

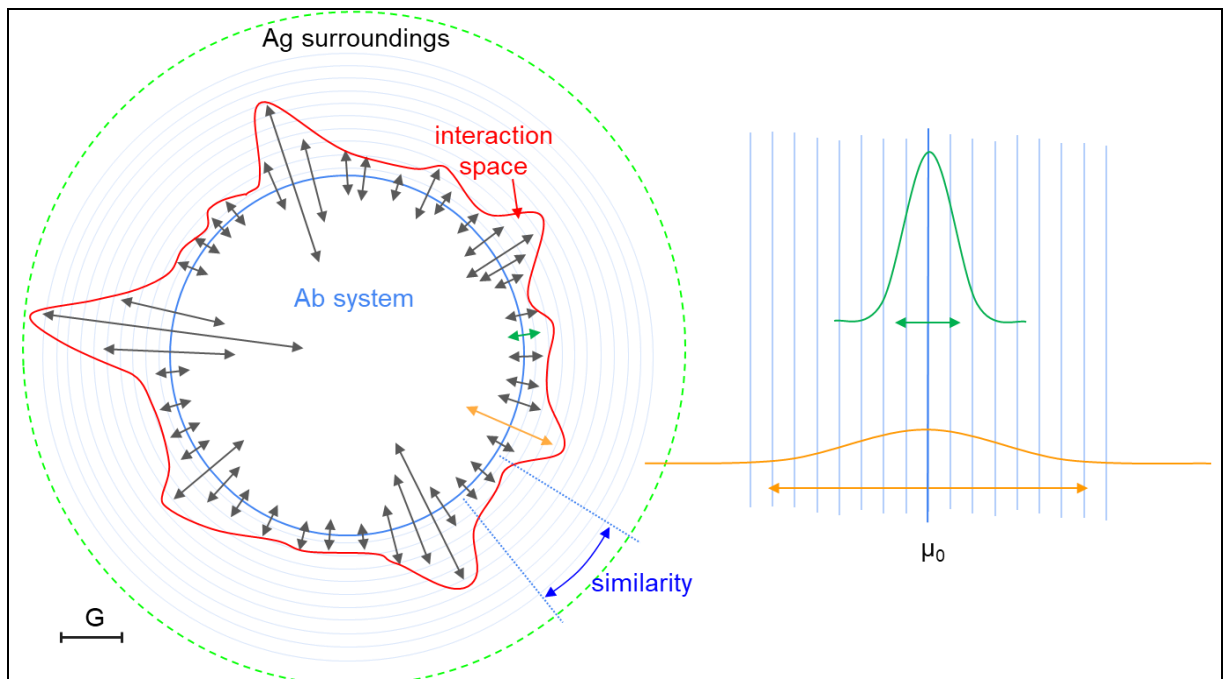

*Figure 3. System organization by mixture of exponential and normal distributions, shown in E scale. Green and yellow double-arrows correspond to the green and yellow normal distributions sampled from the interaction space with exponentially distributed variance.*
*The self-organizing system (Ab system) comprises units (Ab) capable of interacting with the surroundings (Ag surroundings). The free energy of the interactions has normal distribution on a system level. Organization is the adjustment of variance of interaction energy in response to thermodynamic activity of Ag, so as to maintain the system.*

Solving the integral (see online extended data for details), the result is a power law function for $\lambda \leq 1$, that is, for the system core

$$p(\lambda_{Ab}) = \sqrt{2\alpha}\,\lambda_{Ab}^{+\sqrt{2\alpha}-1} \tag{2}$$

whereas for $\lambda \geq 1$, that is, in the interaction space

$$p(\lambda_{Ab}) = \sqrt{2\alpha}\,\lambda_{Ab}^{-\sqrt{2\alpha}-1} \tag{3}$$

These functions are the probability densities of absolute thermodynamic activity of antibody in the system and also define a network of interaction pathways. Thermodynamic activity corresponds to the degree of nodes, in the sense that it represents the probability of interactions. The exponent of network node degree distribution, $-(\sqrt{2\alpha}+1)$, is thus determined by the rate of exponential distribution of chemical potential of antigen molecules, $\alpha$, generating the interaction network.

## 2.4. Architecture of interaction space: a hierarchical, scale-free network

Interaction space spans the range of interaction energies between $\mu_0$ and the upper limit of non-covalent binding energy. This is the space where effector antibodies establish a network of interactions, an energy transduction network for antigen transport. For the description of interaction space we can use another mathematical approach, the combination of exponentials, as suggested by Reed (30) and Newman (31). Reed proposed that when observing exponential growth after an exponentially distributed time, the process will exhibit power law behavior. Here we consider the exponential relationship between antigen chemical potential and absolute thermodynamic activity of antibody, combined with the exponential distribution of chemical potential of antigen. In terms of molecular interactions, the former is the relationship between free energy of the molecule and the number of different states of binding. Since a greater surface area or more non-covalent bonds available for binding (greater buried surface area in the bound molecule) corresponds to greater free energy, absolute thermodynamic activity is a summation of the number of ways of interactions. For a given molecule therefore it is the number of links to other molecular structures with shared binding properties, an expression of similarity in reactivity. The combination of exponentials with the generation of networks can be illustrated by the overlay of activity maps (Fig.4) and mapping of pathways of antigen flow.

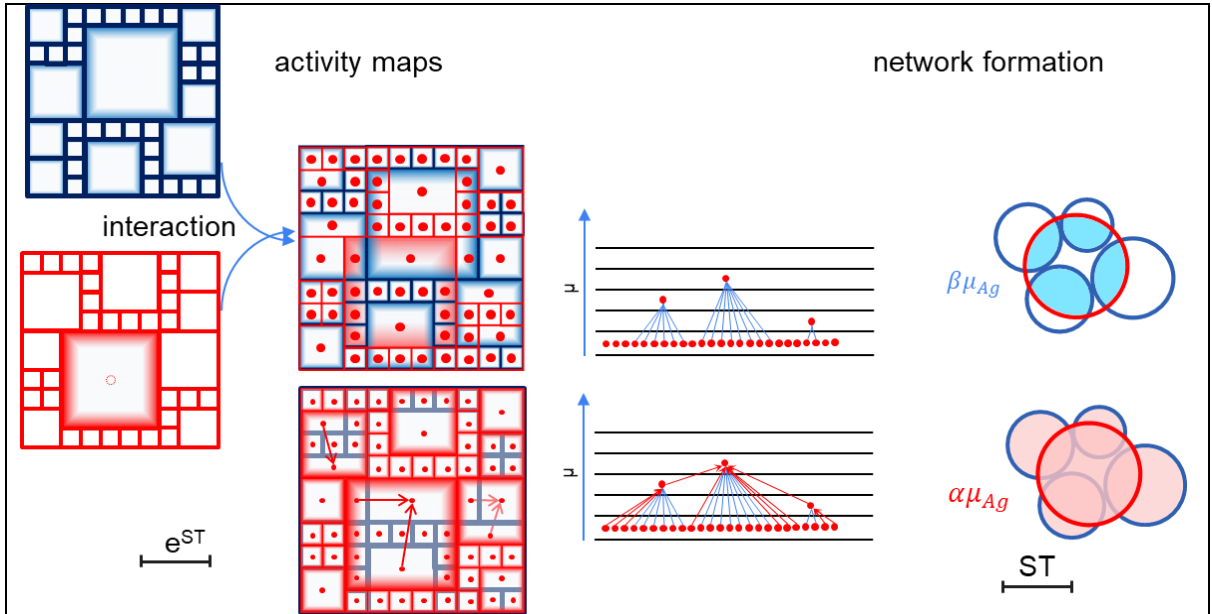

Figure 4. Network hierarchy of molecular interactions.
*Schematic organization of antigen transfer networks. By first overlaying the activity maps of antigens (red grid) and antibodies (blue grid), then renormalizing node connections, we obtain node degree and node weight distributions. Nodes (red dots) are assigned to each box in the overlayed maps. Nodes are sequentially linked within each red box by applying grids with increasing side length and joining nodes in smaller boxes to nodes in the biggest box, until all nodes with the red box are linked. Node hierarchy resulting from own links (blue) and renormalization links (red) are shown in energy diagrams for the shaded red box.*

Again, we assume that the chemical potential μ of antigens is distributed exponentially, when the immune system is in a resting steady state.

$$p(\mu_{Ag}) = \alpha e^{-\alpha \mu_{Ag}} \tag{4}$$

If the absolute thermodynamic activity λ of antibodies is related to Ag chemical potential as

$$\lambda_{Ab} = \beta e^{\beta \mu_{Ag}} \tag{5}$$

then the combination of these exponential functions, the distribution of antigen chemical potential and the thermodynamic activity of antibody is given by

$$p(\lambda_{Ab}) = p(\mu_{Ag})\frac{d\mu_{Ag}}{d\lambda_{Ab}} = \frac{\alpha e^{-\alpha\mu_{Ag}}}{\beta e^{\beta\mu_{Ag}}} = \frac{\alpha}{\beta}\lambda_{Ab}^{-1-\frac{\alpha\mu}{\beta\mu}} = \frac{\alpha}{\beta}\lambda_{Ab}^{-1-\frac{\alpha}{\beta}} \tag{6}$$

In other words, the combination of the exponential distribution of chemical potential of antigens and the exponential relationship between antigen chemical potential and antibody thermodynamic activity determines properties of the energy transduction network, and is defined by the power law distribution of antibody thermodynamic activity. Of note, this expression is an alternative to the system equilibrium binding constant $K_{sys}$ used in a previous publication (19). The value of $\lambda_{Ab}$ reflects the number of pathways a single bond can evolve into a full binding surface and is therefore the number of links to nodes lower in hierarchy. Absolute thermodynamic activity thus corresponds to node degree and its distribution determines network node degree distribution in the system.

Beyond relating Ab and Ag thermodynamic properties, the value of $\beta$ influences how nodes are linked during renormalization. Increasing $\alpha/\beta$ is accompanied by an increasing hierarchy of sub-nodes (Fig.4). From equation 6) we can see that in order to keep the value of degree exponent between 2 and 3, the scale-free network regimen, $\alpha/\beta$ should be in the range $1<\alpha/\beta<2$. Comparing equation 6) and 3) we can see that

$$p(\lambda_{Ab}) = \sqrt{2\alpha}\lambda_{Ab}^{-\sqrt{2\alpha}-1} = \frac{\alpha}{\beta}\lambda_{Ab}^{-1-\frac{\alpha}{\beta}} \tag{7}$$

therefore $\beta = \sqrt{\alpha/2}$ and

$$\frac{\alpha}{\beta} = \sqrt{2\alpha} \tag{8}$$

meaning that the assumption of normally distributed energies with a variance of 1 restricts the possible values of $\beta$, and the value of $\alpha$ alone, via determining the distribution of Ag chemical potentials, determines the distribution of Ab thermodynamic activity and the degree distribution of antigen transport network.

## 2.5. Distribution of energy by links in the system

The humoral adaptive immune system is a transport system for antigen molecules. The flow of antigen molecules in the interaction space of the system, resulting from a concatenated series of interactions, is an energy transduction process. As an antigen molecule is transferred from a lower affinity antibody to a higher affinity antibody, free energy of the system decreases. We may examine in the system the pathways of energy transduction by assigning direction and weight to links. The direction of antigen transport is from lower to higher chemical potential, so links are always directed to nodes with higher degree. We can assign to each link the chemical potential difference between the two nodes connected by the link. The molecular basis of this assignment is the capability of antibodies to take over antigens from weaker binders (lower chemical potential). As figure 4 suggests, renormalization results in the joining of links with lower weight to higher energy nodes, with hierarchy being inversely related to link weight. The number of links assigned to a chemical potential energy level is the sum of in-degrees of the nodes of that level. The average number of interchangeable links to nodes of given chemical potential can be obtained from the relative number of direct links $\beta$, and the rate variable of distribution of antigen chemical potentials $\alpha$, as

$$g_\mu = \frac{\beta}{\alpha} e^{\frac{\beta}{\alpha}\mu_{Ab}} \tag{9}$$

This value is the degeneracy $g$ of the energy level, links of identical chemical potential are interchangeable, represent degeneracy in binding states. The total number of accessible states or molecular partition function $Z$ is given by the integral of the product of (Eq.9) and the exponential distribution of antibody chemical potential obtained from the distribution of thermodynamic activity in (Eq.6), as

$$Z = \int g_\mu p(\mu_{Ab}) \, d\mu = \int e^{-(\frac{\alpha}{\beta} - \frac{\beta}{\alpha})\mu_{Ab}} \, d\mu \tag{10}$$

In other words, with increasing node chemical potential, the increase in activity of the nodes is limited by the number and weight of incoming links, as determined by hierarchy of sharing interaction energies. The logarithm of weighted probabilities of interactions in the energy shells, of the number of microstates the system ranges over, is a thermodynamic potential, entropy. For a system consisting of N elements the entropy is given by

$$S = \frac{U}{T} + N k_B \ln Z \tag{11}$$

An ideal state of the system is described by $\frac{\beta}{\alpha}\mu - \frac{\alpha}{\beta}\mu = -\mu$, that is, when $\frac{\alpha}{\beta} = 1.618...$, the value of golden cut (19). The number of accessible states as a function of chemical potential in this configuration is given by

$$z_\mu = e^{-\mu_{Ab}} \tag{12}$$

and describes energy transport in an ideal stationary state, supported by a golden network, with an optimal balance between increasing node energy and sharing of transport.

### 2.6 Thermodynamic validation of the model

The mathematical constructs obtained from the model so far can be corroborated by understanding the physical meaning of the variables α and β. *Zheng and Wang* proposed (27) that the distribution of equilibrium binding constant is related to the conformational space available at the given temperature and thermal fluctuation, which is in turn related to the heat capacity and flexibility of the molecules (32–34). A more flexible interaction will allow for binding of Ag to distinct Ab clonotypes, exploring greater conformation surface, while a rigid Ag epitope will show stronger preference for a single clonotype. On the other hand, distinct flexible Ab clonotypes may bind to the same Ag, their probability weighted chemical potentials determining average binding energy of Ag. A higher value of $\alpha/\beta$ thus corresponds to greater flexibility and the tail of the distribution becomes flatter: the probability of high affinity interactions in steady state decreases (Fig.5).

In analogy to the heat capacity ratio, which relates heat capacity at constant pressure to heat capacity at constant volume

$$\frac{c_P}{c_V} = \frac{\left(\frac{\partial U}{\partial T}\right)_P}{\left(\frac{\partial U}{\partial T}\right)_V} = \frac{T\left(\frac{\partial S}{\partial T}\right)_P}{T\left(\frac{\partial S}{\partial T}\right)_V} \tag{13}$$

the ratio of $\alpha/\beta$ defines a flexibility ratio in configuration space

$$\frac{\alpha}{\beta} = \frac{T\left(\frac{\partial S}{\partial T}\right)_\mu}{T\left(\frac{\partial S}{\partial T}\right)_N} \tag{14}$$

that relates flexibility with chemical potential $\mu$ kept constant to flexibility with number of system elements $N$ kept constant, in the following sense. The first partial differential expression describes the extent of conformational surface explored by Ag molecules when these molecules are provided by the surroundings so as to keep chemical potential constant. The second expression describes the potential to explore conformational surface when no molecules are supplied, N is constant. The immune system is indeed capable of fine thermodynamic tuning of the interactions of its antibodies by selecting the structure with appropriate flexibility by isotype switching and affinity maturation, and by targeting selected Ag epitopes. The regulation of these Ab properties for each and every Ag is the essence of controlling Ag concentrations.

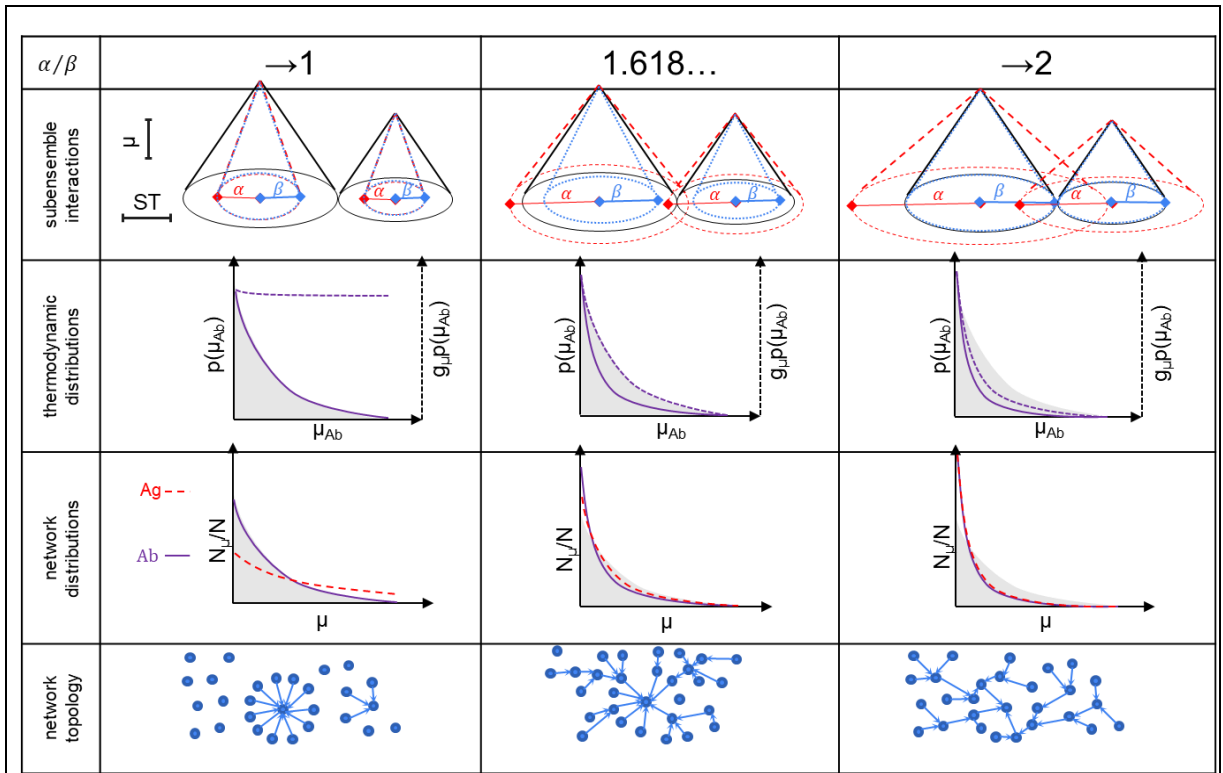

*Figure 5. System properties as determined by the flexibility ratio.*
*The arrangement of binding ensembles in interaction space, the distribution of chemical potentials, partition function and network nodes, and a schematic view of networks is shown for three different values of the flexibility ratio.*

If we regard blood plasma as an open, single-phase, multicomponent system, we can use a thermodynamic potential for the description of energies related to Ab-Ag interactions. Following the nomenclature used by Emmerich (35) for a thermodynamic potential expressed for extensive entropy

$$\Lambda = S + \sum_i \frac{\mu_i}{T} N_i \tag{15}$$

thermodynamic potential $\Lambda$ summarizes the essence of the adaptive thermodynamic system: create a large conformational landscape (conformational entropy S) and generate $N$ elements with chemical potential $\mu$ expressed as

$$\frac{\partial S}{\partial N} = -\frac{\mu}{T} \tag{16}$$

If this system is large enough, we can assume continuity in its distribution and use integrals in the form

$$\Lambda = N\left(k_B \ln Z + \int \frac{\mu}{T} p\left(\frac{\mu}{T}\right) d\mu\right) \tag{17}$$

The configuration space model introduced here arranges constituents of the system based on their "power" to change other constituent's chemical potential. This power is reflected in the topology and network linkage of constituents. In a stationary state the chemical potential of all antigen molecules is kept constant, corresponding to the energy that tickles but not activates sensors, memory or naïve B cells. This is achieved by setting a chemical potential that immunologically suits the host: low chemical potential for harmless self molecules, high chemical potential for dangerous non-self. Concentrations of targets are then adjusted accordingly, meaning that antigens with high chemical potential are eliminated with higher efficiency. The process leads to a wide range of chemical potentials of antigens and antibodies in the system. A theoretical system-wide stationary state is reached when antibody free energies are distributed exponentially.

Self-organization of the system is therefore the distribution of chemical potentials and numbers of constituent entities driven by thermodynamic forces. In a thermodynamic stationary state the system minimizes free energy and maximizes entropy, this state can be described by statistical mechanics functions. The resulting system is an *adaptive, self-organizing biological ensemble*: a co-existing collection of large numbers of copies of states, with a fixed total extensive free energy G and a composition that may fluctuate, with the chemical potential μ distribution adjusted via network formation (Table 1).

*Table 1. Comparison of physical ensembles*
*Pressure is considered constant everywhere. The self-organizing ensemble maintains its extensive Gibbs free energy G by adjusting the chemical potential of its sub-ensembles against elements of the surroundings in interaction space.*

| ensemble | constant | adjusted |
|---|---|---|
| microcanonical | E | microstates |
| canonical | T | E |
| grand canonical | T, μ | E, N |
| adaptive, biological, self-organizing | T, G=∑μN | μ, N |

The biological self-organizing system diversifies its tool of energy transfer, antibody in our case, adjusting its chemical potential so as to maintain the system against forces of change in the environment. By deploying mechanisms to sense free energy of constituents in the environment, the system can adjust, re-adjust and evolve with the environment.

## 2.7 Immunological validation of the model

The humoral immune system must regulate over a very wide range the concentration of a vast number of molecules that are found in and constitute an organism. How it is technically possible to eliminate certain molecules while leave others unharmed leads to the long standing question of tuning recognition specificity and affinity (36), breadth and depth (37). The solution provided by an adaptive immune system apparently requires a scale of numerosity, diversity and affinity comparable to that found in the organism – or more accurately supraorganism (38) - itself.

A mechanistic approach to the feedback and tuning procedure, based on saturation of antibody and antigen, was proposed in a quantitative model of B-cell development and antigen removal (39,40). The model treated distinct molecules independently, neglecting the effect of cross-reactivity, but nevertheless providing a general framework for understanding the system. On the level of individual cells, the level of engagement of antigen receptors determines the cell's fate: programmed death, survival, proliferation and differentiation. An initial repertoire of naive cells, which is constantly generated, displays receptors produced by random rearrangement of gene segments and subsequent selection steps. The immunologically controlled encounters of these cells with the antigens of the inner and outer environment leads to the expansion and differentiation of clones according to the immunological ranking of the target antigen. These processes are going on continuously, manifesting as the adaptation of the system to the antigenic environment, or in other words as self-organization. As a result, a repertoire of memory B cells and long-lived plasma cells are produced, which reflect past adaptation and serve as basis of further evolution in response to the environment (20)(41).

The present model identifies the physical chemical variables that characterize network formation and define the system on a thermodynamic basis. The hierarchic organization of nodes corresponds to the combined action of antibodies with cross-reactive binding properties. It means that multiple B cell clones are producing antibodies against the same epitope, a polyclonal monospecific response. A combination of local antibody diversity and affinity maintains steady state with the antigen. Sensors (memory B cells) are expected to be spread out over the interaction space at positions where local free energy of antigens sustains their survival. Changes in the composition of antigens can trigger the activation of these cells, leading to readjusting properties of the repertoire. It is tempting to speculate that the local hierarchy of binding has important immunological consequences: minimal hierarchy would result in "powerful" antibody response, in the sense of high affinity binding and appropriately switched Ig isotypes, such as an allergic IgE response. With the application of novel tools for the deeper analysis of serological reactions such questions can be experimentally addressed (18).

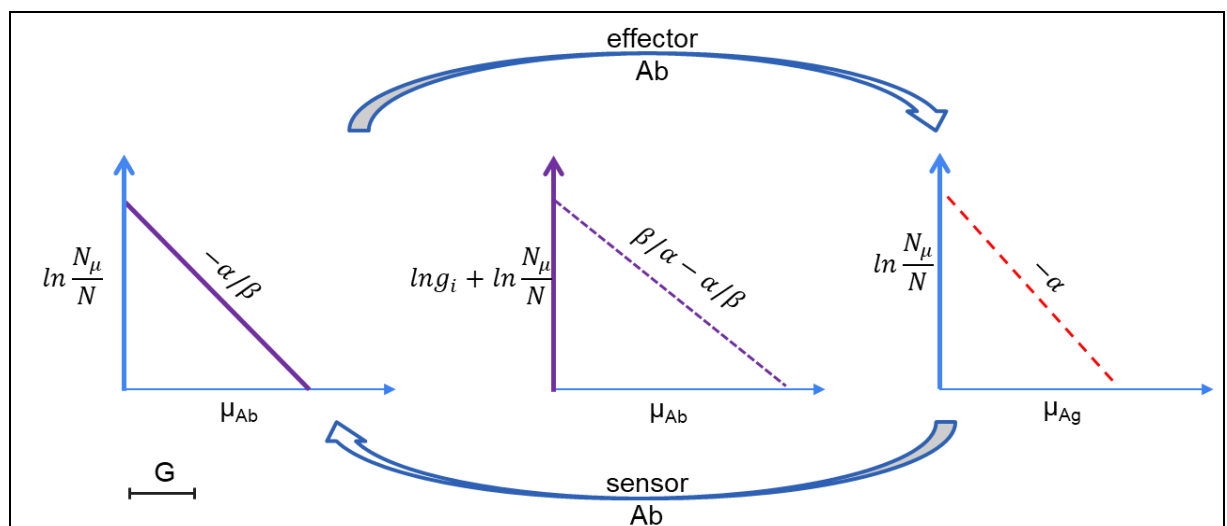

*Figure 6. Immunological aspects of the distributions.*
*The exponential distribution of chemical potentials of both the Ab and Ag are maintained by the feedback loop of effector and sensor Ab. On a logarithmic scale the slope of the distributions are indicated for Ab chemical potentials (solid purple), AbAg complexes (dashed purple line) and Ag (red). The immune system reduces the concentration of Ag according to its chemical potential that is adjusted by Ab and removes AbAg complexes to maintain Ag flow.*

Immunoassays probe the system of antibodies via measuring Ab binding to selected Ag. These assays are also called serological assays because they characterize serum Ab reactivity against medically relevant Ag targets. An immunoassay that uses Ag titration, a gradient of Ag concentrations, can be interpreted as the measurement of the changes of chemical potential in interaction space and can be modeled by the Richards differential equation (20). In the sigmoid Richards curve, which is a generalized logistic function, the point of inflection corresponds to the sum of probability weighted chemical potentials and is therefore a measure of affinity. Meanwhile the asymmetry parameter of the function is related to α/β in the system, which corresponds to the limiting activity coefficient of Ag. Thus, the variable that describes hierarchy and network organization of Ab in the system appears in a biochemical measurement as a thermodynamic variable (18).

A theoretically important message of the model is the continuity of self. The interaction space belongs to the system, it is self. But it also incorporates elements of the surroundings in a regulated way. Antigen binding energies in the interaction space range from system average to very far from average, and elements are distributed according to an exponential distribution. Therefore, the frequency or mole fraction contribution to the system also covers a very wide range (Fig.5). All binding events belong to the system, becoming incorporated into the architecture as imprints in the network. This is consistent with liquid hypothesis of self, which states that immunological identity is continuous and dynamic (42).

Even though we have no exact physical models describing the functioning of the immune system, we have more and more experimental data on the structural and molecular mechanisms of Ab function and networks of Ab based on sequencing. It is thus possible to relate our model to those observations. The total immunoglobulin concentration ($[Ab]_t$) in adult human plasma is in the high micromolar range (~$10^{-4}$ M). We can express Ab quantities as a fraction of the system by using this reference as $[Ab]/[Ab]_t$ (Fig.7).

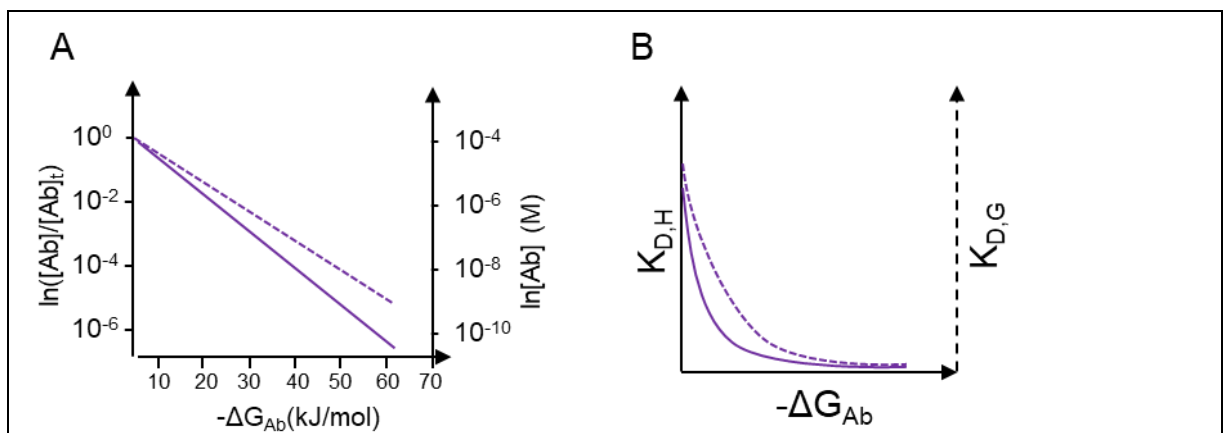

*Figure 7. Antibody concentrations and equilibrium affinity constants. Solid purple line represents values calculated from enthalpy, dashed line represents calculations with free energy.*

Assuming that the system is in a steady state when all its components are present at their equilibrium dissociation concentrations $K_D$, their distribution as a function of chemical potential is determined by the standard free energy of their interactions. This constant has two components

$$K_D = e^{\frac{\Delta_r G°}{RT}} = e^{-\frac{\Delta_r S°T}{RT}} * e^{\frac{\Delta_r H°}{RT}} \qquad (18)$$

namely an entropic and an enthalpic component. The difference between using only the enthalpic component or both components, which corresponds to calculating with p(µ) and $g_\mu p(\mu)$, respectively, is illustrated by the solid and dashed lines of Figure 7. The concentration of an Ab clone with a particular binding energy in the system in stationary state is higher than that of the same Ab in equilibrium with its single binding partner.

**2.8. Network interpretation**

A theoretical network of antibodies can be generated on the basis of sequence and structure similarity (19), which corresponds to experimentally determined, antigen-binding correlation networks (43). On the molecular level these networks allow transfer of antigen from antibody node to another antibody node, in the direction of increasing affinity. The configuration space model of this article organizes such links into a hierarchy of chemical potential and renders structural similarity into topological relationships. Increasing chemical potential in the configuration space means shifting in the hierarchy and increasing node degree, which is related to the probability of receiving an antigen molecule from a node lower in hierarchy. All nodes high in the hierarchy are "supported" by a large number of nodes with lower chemical potential. This arrangement causes the spreading of hubs over the interaction space and manifests as repulsion of hubs and results in a disassortative network.

The power law degree distribution and scale-independent property of the network was previously derived from a lognormal distribution of equilibrium binding constants of random interactions in combination with an exponential distribution of standard deviation of binding energy (19). Here a power law distribution is also obtained from a Boltzmannian distribution of molar free energy of antigen and the relationship between absolute thermodynamic activity of antibody and chemical potential of antigen. In both cases the reference unit of energy is the standard deviation of binding energy, which is the minimal free energy change detected by the biological unit of the system, the B cell. It seems appropriate that levels of hierarchy need to be differentiable by the organizing entities of the system. Importantly, the same units characterize structural differences or vector directions, since cells equipped with antibody sensors recognize antigen structure and quantify structural similarity, as well. Cross-reactivity is the result of partial interactions with lower than maximal energy. Thus, antibodies and target antigens can engage in interactions with partners not identified by the vector but close to the vector. In this case the energy of the interaction decreases with the distance from the vector due to the smaller buried surface area and lower number of non-covalent bonds involved in binding. Higher energy interactions allow more cross-reactivity, since a greater molecular surface area is available for binding. This cross-reactivity appears in the renormalization strategy (Fig.4) where nodes with lower degree are joined to nodes with a higher degree. This renormalization is similar to the box counting method of determining fractal dimension (44), in the sense that grids with increasing side lengths are applied to reveal network architecture. Disassortativity and fractality has been recognized as a feature of biological networks (45,46).

From the network perspective, the ratio of our variables $\alpha/\beta$ corresponds to the ratio of fractal dimension $d_B$ and degree exponent of boxes $d_k$, as described by *Song et al.* (45,47). The network degree distribution exponent $\gamma$ can be calculated from these as

$$\gamma = 1 + \frac{d_B}{d_k} = 1 + \frac{\alpha}{\beta} \tag{19}$$

This is in agreement with our findings that $\alpha/\beta$ is the rate parameter of the exponential distribution of chemical potential and $\frac{\alpha}{\beta} + 1$ is the exponent of the power law distribution of thermodynamic activity and network degree distribution (Eq.6)(also see Supplementary data). The evolution of the network in time is reflected in configuration space, in as much as the nodes with greater chemical potential that develop in the time course of immunological reactions are located more distantly from the system surface. Therefore these indices can be interpreted as factors of renormalization in energy levels.

According to Caetano-Anollés et al. (48) scale-free networks can follow a trajectory in network morphospace from homogenous, non-modular towards heterogenous modular network organization. This trajectory can correspond to a mesh-like structure of random binding events at the core surface of the system evolving into a highly hierarchical and modular network of specific and high affinity interactions. This corresponds to a star-like network distribution of energy when α/β approaches 1, and multiple transfers of energy when α/β increases (Fig.5).

Whereas the probability density of energy is associated with a Boltzmannian exponential distribution, scale-free networks are known for the power-law distribution of network node degree. Our configuration space model of a system suggests that these two phenomena are two sides of the same coin: an identical complementary cumulative distribution results in the exponential probability density function of chemical potential and power law probability function of node degrees (see online extended data) of elements of the system.

**2.9. Complex systems interpretation**

In their maximum entropy model for Ab diversity *Mora et al.* associate Ab sequences with effective energies as if the sequences represented a particular state in a system in equilibrium (16). In this article it is suggested that antibody structures, which are defined by their amino acid sequences, are indeed distinct states in the configuration space of the system. The molecular free energy of these structures is determined by the composition of the system, which is itself regulated by the system.

Following the active, expansive stage of the immune response a contraction stage establishes a balance between the thermodynamic activity of the antibody and of the antigen by maintaining only a subset of the cells that evolved in the active stage. This process leads to a steady state, whereby secreted antibodies are continuously removed along with bound antigen, keeping both free antibody and free antigen concentrations at the immunologically adjusted values. A two-way feedback mechanism of sensor and effector cells adjusts chemical potential so that cells are poised between under- and overactivation (see (8,20,39) for details). The effector cell secretes antibodies capable of reducing target antigen concentration, the extent of reduction is determined by chemical potential of antibody. The sensor cell receives an amount of energy in the form of survival signals by target antigen – this is called tickling in immunological jargon. This signal is adjusted by the effector cells, secreting soluble antibodies. From the immunological point of view, this means the parallel generation of memory B cells (sensors) and plasma cells (effectors) in germinal centers with closely related if not identical binding properties (49–51). Perturbations at any point in interaction space trigger the rearrangement of this hierarchy and network. In this sense the system is poised at criticality, a phenomenon suggested to be present in all biological systems (52). Critical events represent the reorganization of hierarchy in a system. These events are the coalescence of sub-ensembles or subnetworks. The size of these events often shows power law distribution over time (strength of earthquakes, size of forest fires, avalanches of sandpiles) (53). The immune system presumably reorganizes itself constantly, adapting to the antigenic environment, via such events. Occasionally, triggered by infections, massive reorganization is necessary, which may correspond to a huge critical event.

A variety of phenomena have been shown to follow power law, including natural events and systems and artificial, man-made systems. In these systems the power law applies only for values greater than some minimum value $x_{min}$. The distributions described here suggest that the minimum value of x corresponds to activity with the average energy or reference chemical potential $\mu_0$ of the system.

$$\lambda_{min} = e^{\mu_0} = 1 \tag{20}$$

By our definition this reference value is zero and the corresponding activity is one. It also means that in our directed network all isolated nodes possess an in-degree of 1. Nodes with higher degree connect to form the networks, which represent the organization of interactions of the system with its environment.

Power law relationships can be broadly assigned into two main categories: distributions of frequencies and of magnitudes (54). Complementary cumulative distributions (cCDF) of frequencies, rank-frequency plots, the Zipf-law, the Pareto distributions are examples of the first type (29,31). Protein interaction and metabolic networks are examples for the second type (55,56).

Magnitudes represent an intensive physical property of the adaptive system. The distribution of an intensive physical property describes a hierarchy and often represents a hierarchical network that organizes the system. It follows those magnitudes are related to the network degree of the entities. The configuration space model presented here suggests that those magnitudes are absolute thermodynamic activities and are related to molar free energy or chemical potential, an intensive physical property.

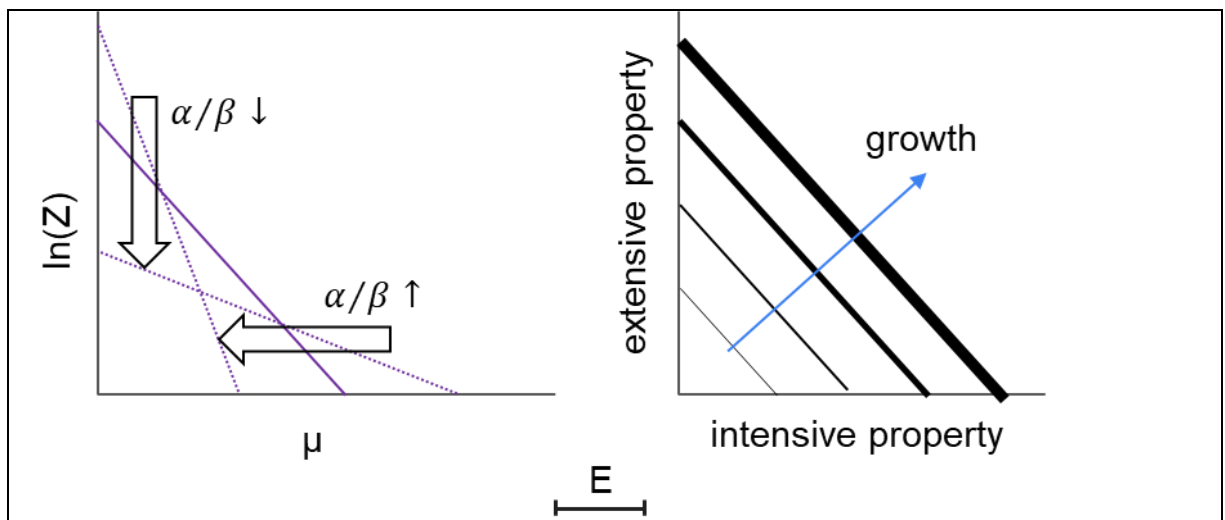

*Figure 8. Graphical representation of relationships of system variables.*
*The relationship between entropy and chemical potential of system elements reults from the balanced distribution of interaction energies and accessible binding states. Balanced growth maintains organization by maintaining relations between intensive and extensive properties.*

Frequency is an extensive property and corresponds to the number of links (interactions) of a given energy in our model. If the energy of interactions is distributed exponentially (Eq.8), then the logarithm of probability (Eq.9) decreases linearly (Fig.6). The corresponding linear growth of entropy (for cumulative distribution) is characteristic of criticality in thermodynamic systems (16,57). In an unchanging, simple surroundings the system can strongly adapt (Fig.6, vertical arrow) and distribute its energies over a wider range of chemical potentials. On the other hand, a changing, diverse surroundings prohibit effective adaptation (Fig.6, horizontal arrow) allowing large numbers of weak interactions. The system can develop and grow maintaining its organization by parallel growth of its intensive and extensive properties, chemical potentials and the number of interactions as a function of those potentials, respectively.

Power law distributions in general characterize critical points and phase transition. In this respect, self-organization aims to maintain the state of phase transition. In an abstract sense, transition is between the two phases of organization of matter in the system and in the surroundings. Self-organization thus counteracts the effects of the surrounding environment by generating an interaction space and therein sustaining phase transition, maintaining steady state reactions towards all environmental components that would otherwise disintegrate the system.

## 3. Conclusions

There can be little doubt that a system with the number of constituents and the extent of diversity that the adaptive immune system possesses could behave as a complex adaptive physical system. The more important question is perhaps: would a complex biological system follow the rules that apply to a thermodynamic system? This article identifies attributes of the humoral adaptive immune system that seem to capture the physics of the biological system. These attributes are magnitudes and frequencies of binding energies, chemical potentials and their weighted probabilities, respectively. A coefficient that appears in all approaches of description of the system is the heat capacity ratio, or flexibility ratio or degree distribution exponent. Beyond the theoretical advancement in modeling, the introduced mathematical framework can also be put into practice in quantitative serological measurements (18,20), where this coefficient can be experimentally obtained.

**Supplementary Online Material**
Supplemental text file 1 describes the mathematical derivation of power law distribution from the combination of exponential and lognormal distributions. Supplemental text 2 summarizes the relationships between exponential and power law distributions.

**Acknowledgments**
Special thanks to Tamás Pfeil (ELKH-ELTE Numerical Analysis and Large Networks Research Group, Budapest) for reading the manuscript and correcting my mathematical mistakes. Without his help I could not have clarified the math behind these phenomena. Many thanks to Tamás Vicsek (ELTE, Department of Biological Physics, Budapest) for taking his time to read the manuscript and advise me on the backgrounds of this huge field of science.

**Conflict of interest**
The author declares no conflict of interest.

**Funding**
The laboratory of J.P. is partly supported by the ENDONANO H2020 Marie Skłodowska-Curie ITN programme.